\documentclass[prb,aps,twocolumn]{revtex4-1}
\usepackage{graphics, bm, psfrag, amsmath, amssymb, epsfig, grffile, float, bbold}
\usepackage{subfigure}
\usepackage{dsfont}
\usepackage{color}
\usepackage{hyperref}

\begin{document}

\title{Odd-frequency superconductivity induced by non-magnetic impurities}
\author{Christopher Triola}
\affiliation{Department of Physics and Astronomy, Uppsala University, Box 516, S-751 20 Uppsala, Sweden}
\author{Annica M. Black-Schaffer}
\affiliation{Department of Physics and Astronomy, Uppsala University, Box 516, S-751 20 Uppsala, Sweden}
%-------------------------------------------------------
%-------------------------------------------------------

\begin{abstract}
A growing body of literature suggests that odd-frequency superconducting pair amplitudes can be generated in normal metal-superconductor junctions. The emergence of odd-frequency pairing in these systems is often attributed to the breaking of translation invariance. In this work, we study the pair symmetry of a one-dimensional $s$-wave superconductor in the presence of a single non-magnetic impurity and demonstrate that translation symmetry breaking is not sufficient for inducing odd-frequency pairing. We consider three kinds of impurities: a local perturbation of the chemical potential, an impurity possessing a quantum energy level, and a local perturbation of the superconducting gap. Surprisingly, we find local perturbations of the chemical potential do not induce any odd-frequency pairing, despite the fact that they break translation invariance. Moreover, although odd-frequency can be induced by both the quantum impurity and the perturbation of the gap, we find these odd-frequency amplitudes emerge from entirely different kinds of scattering processes. The quantum impurity generates odd-frequency pairs by allowing one of the quasiparticles belonging to an equal-time Cooper pair to tunnel onto the impurity state and then back to the superconductor, giving rise to odd-frequency amplitudes with a temporal broadening inversely proportional to the energy level of the impurity. In contrast to this, the perturbation of the gap leads to odd-frequency pairing by ``gluing-together" normal state quasiparticles from different points in space and time, leading to odd-frequency amplitudes which are very localized in the time domain.      
\end{abstract}

%-------------------------------------------------------

\maketitle

\section{Introduction}

It is widely appreciated that the symmetries of a many-body wavefunction are tightly constrained by the statistics of the constituent particles. In the case of superconductors, the fermionic nature of the quasiparticles comprising the Cooper pairs implies that the superconducting gap function must be odd under the exchange of all the quantum numbers of the Cooper pair. This usually limits the pairing to either spatially even-parity gap functions (like $s$- or $d$-wave) with spin-singlet configuration or odd-parity gap functions ($p$- or $f$- wave) corresponding to spin-triplet states. However, if the quasiparticles pair at unequal times the superconducting gap can also be odd in time or, equivalently, odd in frequency (odd-$\omega$), thus allowing for Cooper pairs that are even in spatial parity and  spin-triplet or odd-parity and spin-singlet.\cite{bergeret2005odd,linder2017odd} 

First discussed by Berezinskii\cite{Berezinskii1974} in the context of $^3$He, and later generalized to superconductivity,\cite{kirkpatrick_1991_prl,belitz_1992_prb,BalatskyPRB1992} odd-$\omega$ pairing is intriguing both because of the unconventional symmetries which it permits and for the fact that it represents a class of hidden order, due to the vanishing of equal-time correlation functions. The possibility of odd-$\omega$ pairing can readily be seen by inspecting the anomalous Green's function, $F_{\sigma_1,x_1,\alpha_1,t_1;\sigma_2,x_2,\alpha_2,t_2}$, with indices $\sigma_i$, $x_i$, $\alpha_i$, and $t_i$ labeling the spin, position, orbital, and time degrees of freedom of the Cooper pairs, where the orbital index could also be interpreted as a sublattice or band label. Accounting for the symmetries of $F$ under the permutation of these four indices, all Cooper pair amplitudes must fall into one of eight possible classes: four even-$\omega$ classes and four odd-$\omega$ (see Table~\ref{table:classification}). 

While the thermodynamic stability of intrinsically odd-$\omega$ phases has, so far, only been discussed as a theoretical possibility,\cite{coleman_1993_prl,coleman_1994_prb,coleman_1995_prl,heid1995thermodynamic,belitz_1999_prb,solenov2009thermodynamical,kusunose2011puzzle,FominovPRB2015} significant progress has been made toward understanding the way in which odd-$\omega$ pairing can be induced by altering a system's conventional superconducting correlations through symmetry breaking.\cite{BergeretPRL2001, bergeret2005odd, halterman2007odd, yokoyama2007manifestation, houzet2008ferromagnetic, EschrigNat2008, LinderPRB2008, crepin2015odd, YokoyamaPRB2012, Black-SchafferPRB2012, Black-SchafferPRB2013, TriolaPRB2014, tanaka2007theory, TanakaPRB2007,cayao2017odd, cayao2018odd, LinderPRL2009, LinderPRB2010_2, TanakaJPSJ2012, triola2016prl, triolaprb2016, 
weiss2017odd,
sothmann2014unconventional,burset2016all,
ebisu2016theory,cayao2018odd,triola2018oddnw,
kuzmanovski2017multiple,cayao2017odd,keidel2018tunable,fleckenstein2018conductance, 
black2013odd, parhizgar_2014_prb,  asano2015odd, komendova2015experimentally, komendova2017odd, triola2017pair, triola2018odd,asano2018green} The best established example is found in superconductor-ferromagnet (SF) junctions,\cite{BergeretPRL2001, bergeret2005odd, halterman2007odd, yokoyama2007manifestation, houzet2008ferromagnetic, EschrigNat2008, LinderPRB2008, crepin2015odd} in which experiments have observed key signatures of odd-$\omega$ spin-triplet pair correlations,\cite{zhu2010angular,di2015signature,di2015intrinsic} despite using conventional spin-singlet $s$-wave superconductors. The key property here is that both the interface and the bulk magnet break the symmetry between up and down spins, thus allowing for the conversion of spin-singlet to spin-triplet $s$-wave Cooper pairs at the interface, where the latter is necessarily odd in frequency. 

\begin{center}
\begin{table}[htb]
\begin{tabular}{c || c | c | c | c || c | c | c | c |}
        & 1 & 2 & 3 & 4 & 5 & 6 & 7 & 8 \\
\hline 
Spin ($\mathcal{S}$ )   & - & + & + & - & + & - & - & +  \\ 
\hline
Parity ($\mathcal{P}$)  & + & - & + & - & + & - & + & -  \\
\hline
Orbital ($\mathcal{O}$) & + & + & - & - & + & + & - & -  \\
\hline
 Time  ($\mathcal{T}$)  & + & + & + & + & - & - & - & -  \\
\end{tabular}
\caption{Characterization of the eight symmetry classes for the superconducting gap function allowed by Fermi-Dirac statistics. Each column represents a different symmetry class, with the sign, $\pm$, representing the symmetry of the anomalous Green's function under the exchange of the index indicated in the far left column: $\mathcal{S}F_{\sigma_1,\sigma_2}= F_{\sigma_2,\sigma_1}$ (spin); $\mathcal{P}F_{x_1,x_2}=F_{x_2,x_1}$ (parity); $\mathcal{O}F_{\alpha_1,\alpha_2}=F_{\alpha_2,\alpha_1}$ (orbital); and $\mathcal{T}F_{t_1,t_2}=F_{t_2,t_1}$ (time).}
\label{table:classification}
\end{table}
\end{center}

Symmetry-breaking has also been employed in proposals for finding odd-$\omega$ odd-parity pairing at the interface between a normal metal and a conventional superconductor in SN junctions. Here the interface breaks spatial translational invariance, resulting in a transformation of even-$\omega$ $s$-wave spin-singlet pairing to odd-$\omega$ odd-parity spin-singlet pairing.\cite{tanaka2007theory, TanakaPRB2007} Similar strategies have also led to predictions of odd-$\omega$ pairing in mesoscopic superconducting systems including single quantum dots in the presence of magnetic fields,\cite{weiss2017odd} double quantum dots,\cite{sothmann2014unconventional,burset2016all} nanowires,\cite{ebisu2016theory,cayao2018odd,triola2018oddnw} and the edge modes of two-dimensional topological insulators.\cite{kuzmanovski2017multiple,cayao2017odd,keidel2018tunable,fleckenstein2018conductance} Moreover, analogous methods have been used to demonstrate that odd-$\omega$ pairing should be ubiquitous in multiorbital/multiband systems when the orbitals/bands are coupled and there is an imbalance, or symmetry breaking, between them.\cite{black2013odd, parhizgar_2014_prb,  asano2015odd, komendova2015experimentally, komendova2017odd, triola2017pair, triola2018odd,asano2018green}

Surveying the literature, one therefore gets the impression that odd-$\omega$ pairing can be generated by: (i) starting with a superconductor belonging to one of the even-$\omega$ symmetry classes in Table~\ref{table:classification}; and (ii) applying an external field which breaks any one of the symmetries of the Cooper pairs, i.e~spin ($\mathcal{S}$), parity ($\mathcal{P}$), orbital ($\mathcal{O}$), or time ($\mathcal{T}$). The resulting odd-$\omega$ pair amplitudes should then be proportional to the symmetry-altering field. In this work we highlight a particularly simple setup in which these steps demonstratively fail to generate odd-$\omega$ pairing. By examining the physical reasons for this failure we gain important insight into the dynamical nature of odd-$\omega$ Cooper pair formation that is crucial for understanding its emergence. 

Specifically, we consider a uniform one-dimensional (1D) $s$-wave spin-singlet superconductor in the presence of a single non-magnetic impurity, which manifestly breaks the translation invariance of the superconductor. We model the impurity in three different ways: as a classical potential impurity only modifying the chemical potential of the quasiparticles at point $x_0$; as a quantum impurity localized at point $x_0$ with an on-site energy level, $\epsilon_0$, coupled to the superconductor with a tunneling amplitude, $t_0$; and as a local anomalous impurity modifying the superconducting gap at point $x_0$. Surprisingly, we find that odd-$\omega$ pairing cannot be generated by classical potential impurities, despite the fact that these impurities manifestly break translation invariance in the superconductor. Moreover, we find that this conclusion holds even in the presence of an arbitrary number of classical impurities. However, we do find that odd-$\omega$ pairing can be generated by both the quantum impurity and the anomalous impurity, but through very different mechanisms. 

In the case of the quantum impurity, we find that odd-$\omega$ pairs are created by processes in which one of the two quasiparticles making up an equal-time Cooper pair tunnels to the impurity site and then back to the superconductor. In this case, it is, thus, the finite lifetime of the impurity state that is crucial for the formation of odd-$\omega$ Cooper pairs, which leads to the odd-$\omega$ pair amplitude being quite broad in the time domain. In contrast to this, the anomalous impurity induces odd-$\omega$ pairing through processes involving two normal quasiparticles in the superconductor from different points in space, $x_1$ and $x_2$, scattering at the site of the locally modified gap, $x_0$. This mechanism gives rise to odd-$\omega$ pair amplitudes which, in the time domain, are highly localized at $t=(|x_1-x_0|+|x_2-x_0|)/v_F$, where $v_F$ is the fermi velocity in the superconductor. Taken together, our results show that odd-$\omega$ pairing can be generated by two very different mechanisms in non-magnetic systems, which not only highlights the dynamical nature of this pairing but also explicitly demonstrates that translation symmetry breaking is not sufficient for the generation of odd-$\omega$ pair amplitudes. Furthermore, by examining the physical origins of odd-$\omega$ pair amplitudes, our results should be useful for designing superconducting systems with more odd-$\omega$ pairing.

The remainder of this work is organized as follows. In Sec.~\ref{sec:potential_imp} we study a classical potential impurity and show that odd-$\omega$ pairing does not emerge, even when an arbitrary number of impurities are included. In Sec.~\ref{sec:quantum_imp} we turn our attention to a quantum impurity and find analytical expressions for the induced odd-$\omega$ pairing. In Sec.~\ref{sec:anomalous_imp} we consider an impurity which acts as a local modification of the superconducting gap, and derive expressions for the odd-$\omega$ pair amplitudes. In Sec.~\ref{sec:time} we compare the odd-$\omega$ pair amplitudes obtained for the quantum and anomalous impurities by inspecting their behavior in the time domain and find qualitative differences related to the disparate physical origins of the odd-$\omega$ pairing in the two cases. Finally, in Sec.~\ref{sec:conclusions} we offer our conclusions.   

\section{Pair Symmetry in the presence of a potential impurity}
\label{sec:potential_imp}
We wish to study the emergent symmetries of Cooper pair amplitudes in a conventional spin-singlet $s$-wave superconductor in the presence of a particularly simple instance of translational invariance breaking: a single non-magnetic impurity. To simplify the calculations in real space we restrict ourselves to a 1D superconductor using a model which effectively captures the low-energy physics of a 1D nanowire proximity-coupled to a superconducting substrate.\cite{cayao2018odd} However, it is important to emphasize that the generalization of our main results to higher dimensions is very straightforward. 

We begin by considering the simplest non-magnetic impurity, the classical potential impurity, which can be understood as a local perturbation of the chemical potential of the superconductor, modeled by the Hamiltonian 
\begin{equation}
H=H_{\text{SC}}+H_{\delta\mu}
\label{eq:ham_pot}
\end{equation}
where $H_{\text{SC}}$ is the Hamiltonian describing a clean 1D superconductor,
\begin{equation}
\begin{aligned}
H_{\text{SC}}&=\sum_\sigma\int dx \psi^\dagger_{x,\sigma}\left(-\frac{\partial_x^2}{2m}-\mu\right)\psi_{x,\sigma} \\
&+\int dx\Delta \psi^\dagger_{x,\uparrow}\psi^\dagger_{x,\downarrow} + \text{H.c.}, 
\end{aligned}
\label{eq:ham_sc}
\end{equation}
where $\psi^\dagger_{x,\sigma}$ ($\psi_{x,\sigma}$) creates (annihilates) a quasiparticle state in the superconductor with spin $\sigma$ at position $x$. Here $m$ is the effective mass, $\mu$ the chemical potential, and $\Delta$ the superconducting order parameter, assumed to be real and uniform throughout the superconductor. The local perturbation, $H_{\delta\mu}$, is given by 
\begin{equation}
\begin{aligned}
H_{\delta\mu}&=\delta\mu\sum_\sigma\int dx \delta(x-x_0) \psi^\dagger_{x,\sigma}\psi_{x,\sigma}, 
\end{aligned}
\label{eq:ham_pot_imp}
\end{equation}
where $\delta\mu$ is the strength of the classical impurity at position $x_0$.

To study the emergent electronic properties of this system, we begin by defining the normal and anomalous Green's functions for the superconductor in the usual way for a non-magnetic system
\begin{equation}
\begin{aligned}
G_{x_1,x_2}(\tau)&=-\langle T_\tau \psi_{x_1,\uparrow}(\tau)\psi^\dagger_{x_2,\uparrow}(0) \rangle, \\
F_{x_1,x_2}(\tau)&=-\langle T_\tau \psi_{x_1,\uparrow}(\tau)\psi_{x_2,\downarrow}(0) \rangle, 
\end{aligned}
\label{eq:g_f_sc}
\end{equation}
where $\tau$ is an imaginary time and $T_\tau$ is the usual $\tau$-ordering operator for fermions. For convenience we combine these normal and anomalous Green's functions into the Nambu space Green's function for the superconductor
\begin{equation}
\begin{aligned}
\hat{\mathcal{G}}(x_1,x_2;i\omega_n)&=\left(
\begin{array}{cc}
G_{x_1,x_2;i\omega_n} & F_{x_1,x_2;i\omega_n} \\
F_{x_1,x_2;i\omega_n}^* & -G_{x_1,x_2;i\omega_n}^*
\end{array} \right), \\
\end{aligned}
\label{eq:nambu_g_sc}
\end{equation} 
where we have Fourier-transformed to Matsubara frequency, $i\omega_n$. As is well-known, both the retarded and advanced Green's functions can be obtained from the Matsubara Green's function, Eq.~(\ref{eq:nambu_g_sc}), via the prescriptions: $\hat{\mathcal{G}}^{\text{R}}(x_1,x_2;\omega)=\lim_{\eta\rightarrow 0^+}\hat{\mathcal{G}}(x_1,x_2;\omega+i\eta)$ and $\hat{\mathcal{G}}^{\text{A}}(x_1,x_2;\omega)=\lim_{\eta\rightarrow 0^+}\hat{\mathcal{G}}(x_1,x_2;\omega-i\eta)$. Therefore, throughout this work we derive all expressions in terms of the complex frequency, $z$, thus capturing the results for the Matsubara, retarded, and advanced correlators. 

In the absence of any impurities, the bare superconductor is translation-invariant and it is straightforward to derive the analytic expression for its Green's function using Eq.~(\ref{eq:ham_sc}):
\begin{equation}
\begin{aligned}
\hat{\mathcal{G}}_0(x;z)&=\left(z \hat{\tau}_0 + \Delta \hat{\tau}_1\right)g_0(x;z) + \hat{\tau}_3 g_3(x;z), \\
\end{aligned}
\label{gsc_0}
\end{equation}   
where $\hat{\tau}_0$ and $\hat{\tau}_{i=\{1,2,3\}}$ are the $2\times 2$ identity and Pauli matrices in Nambu space, respectively. Here we have defined the following functions:
\begin{equation}
\begin{aligned}
g_0(x;z)&=-\frac{i\zeta }{2v_F\Omega}\left[\frac{e^{i\zeta|x|k_F\sqrt{1+\tfrac{\Omega}{\mu}}}}{\sqrt{1+\tfrac{\Omega}{\mu}}}+\frac{e^{-i\zeta|x|k_F\sqrt{1-\tfrac{\Omega}{\mu}}}}{\sqrt{1-\tfrac{\Omega}{\mu}}} \right], \\
g_3(x;z)&=-\frac{i\zeta }{2v_F}\left[\frac{e^{i\zeta|x|k_F\sqrt{1+\tfrac{\Omega}{\mu}}}}{\sqrt{1+\tfrac{\Omega}{\mu}}}-\frac{e^{-i\zeta|x|k_F\sqrt{1-\tfrac{\Omega}{\mu}}}}{\sqrt{1-\tfrac{\Omega}{\mu}}} \right], \\
\end{aligned}
\end{equation}   
where $\Omega=\sqrt{z^2-\Delta^2}$, $\zeta=\text{sgn}[\text{Im}(\Omega)]$, $k_F=\sqrt{2m\mu}$, and $v_F=k_F/m$. Since all scattering processes are elastic, we suppress the $z$ argument of all functions for the remainder of this work, except where necessary.

In the presence of a classical impurity described by Eq.~(\ref{eq:ham_pot_imp}), we can compute the total Green's function for the superconductor to infinite order in $\delta\mu$ using the Dyson equation
\begin{equation}
\hat{\mathcal{G}}(x_1,x_2)=\hat{\mathcal{G}}_0(x_1-x_2)+\delta\mu\hat{\mathcal{G}}_0(x_1-x_0)\hat{\tau}_3\hat{\mathcal{G}}(x_0,x_2).
\label{eq:dyson_pot}
\end{equation}
It is straightforward to show that the right-hand-side of this recurrence relation can be rewritten as
\begin{equation}
\hat{\mathcal{G}}(x_1,x_2)=\hat{\mathcal{G}}_0(x_1-x_2)+\hat{\mathcal{G}}_0(x_1-x_0)\hat{T}_{\delta\mu}\hat{\mathcal{G}}_0(x_0-x_2),
\label{eq:g_pot}
\end{equation}
where $\hat{T}_{\delta\mu}=\left[\delta\mu^{-1}\hat{\tau}_3-\hat{\mathcal{G}}_0(0)\right]^{-1}$ is a $T$-matrix describing the total effect of the potential impurity. Inserting the bare Green's function expressions from Eq.~(\ref{gsc_0}) we find 
\begin{equation}
\hat{T}_{\delta\mu}=-\frac{\left(z \hat{\tau}_0-\Delta\hat{\tau}_1\right) \delta\mu^2g_0(0) +\delta\mu\left(1-\delta\mu g_3(0) \right)\hat{\tau}_3}{\left(z^2-\Delta^2\right)\delta\mu^2g_0^2(0)-\left(1-\delta\mu g_3(0)\right)^2}. 
\end{equation}
Finally, combining these expressions with Eq.~(\ref{eq:g_pot}), we find that the total Green's function of the superconductor in the presence of a single potential impurity is
\begin{equation}
\begin{aligned}
\hat{\mathcal{G}}(x_1,x_2)&=\left(z \hat{\tau}_0 +\Delta\hat{\tau}_1\right) \left( g_0(x_1-x_2) +\delta g^{(\mu)}_0(x_1,x_2) \right) \\
& + \hat{\tau}_3\left( g_3(x_1-x_2) + \delta g^{(\mu)}_3(x_1,x_2) \right),
\end{aligned}
\end{equation}
where $\delta g^{(\mu)}_0(x_1,x_2)=\delta g^{(\mu)}_0(x_2,x_1)$ and $\delta g^{(\mu)}_3(x_1,x_2)=\delta g^{(\mu)}_3(x_2,x_1)$. Thus, the presence of this potential impurity simply renormalizes the coefficients describing the bare Green's function in Eq.~(\ref{gsc_0}) and, thus, does not generate any odd-$\omega$ pair amplitudes. 
Importantly, the presence of this impurity does not change the spatial parities of any of the coefficients of the Green's function for the superconductor. Using this fact, we can easily iterate the above $T$-matrix calculation to account for $N$ potential impurities and thus show that no amount of potential impurities generate odd-$\omega$ pairing. 

The lack of odd-$\omega$ pairing for a classical impurity is a surprising result, since the literature is full of examples in which conventional superconducting pair amplitudes can be converted to odd-$\omega$ amplitudes by breaking almost any symmetry of the system.\cite{BergeretPRL2001, bergeret2005odd, halterman2007odd, yokoyama2007manifestation, houzet2008ferromagnetic, EschrigNat2008, LinderPRB2008, crepin2015odd, YokoyamaPRB2012, Black-SchafferPRB2012, Black-SchafferPRB2013, TriolaPRB2014, tanaka2007theory, TanakaPRB2007, LinderPRL2009, LinderPRB2010_2, TanakaJPSJ2012, triola2016prl, triolaprb2016, black2013odd, sothmann2014unconventional, parhizgar_2014_prb, asano2015odd, komendova2015experimentally, burset2016all, komendova2017odd, kuzmanovski2017multiple, triola2017pair,keidel2018tunable, triola2018odd,fleckenstein2018conductance,asano2018green} In particular, multiple previous works have found odd-$\omega$ pairing induced in conventional superconductors due to the breaking of spatial translation invariance at SN interfaces.\cite{tanaka2007theory, TanakaPRB2007} In the next section we take a first step toward reconciling this null result with previous research by adding some internal dynamics to the impurity. In this sense, the impurity studied in the next section could be viewed as an infinitesimal normal region in an SN junction. 

\section{Pair symmetry in the presence of a quantum impurity}
\label{sec:quantum_imp}
To better approximate the situation at an SN junction, we next consider the possibility that the local non-magnetic impurity possesses some internal structure. To keep the formalism simple and isolate the key features, we assume a single quantum energy level to which the electrons in the substrate may tunnel, similar to a quantum dot. In this case the Hamiltonian takes the form
\begin{equation}
H=H_{\text{SC}}+H_{\text{imp}}+H_t
\label{eq:ham_anderson}
\end{equation}
where $H_{\text{SC}}$ is the Hamiltonian describing the clean 1D superconductor, given by Eq.~(\ref{eq:ham_sc}), $H_{\text{imp}}$ is the Hamiltonian governing the dynamics of the impurity,
\begin{equation}
\begin{aligned}
H_{\text{imp}}&=\sum_\sigma \epsilon_0 c^\dagger_{\sigma}c_{\sigma}, 
\end{aligned}
\label{eq:ham_and_imp}
\end{equation}
where $c^\dagger_{\sigma}$ ($c_{\sigma}$) creates (annihilates) a fermionic impurity state with spin $\sigma$ and energy $\epsilon_0$, and $H_t$ describes the tunneling between the superconductor and the impurity at position $x_0$,
\begin{equation}
\begin{aligned}
H_{t}&=t_0\sum_\sigma\int dx \delta(x-x_0) \psi^\dagger_{x,\sigma}c_{\sigma} + \text{H.c.}, 
\end{aligned}
\label{eq:ham_t}
\end{equation}
where $t_0$ is the tunneling amplitude.

Again, we solve exactly for the Green's function in the superconductor using the $T$-matrix formula in Eq.~(\ref{eq:g_pot}) but now with the $T$-matrix given by $\hat{T}_{Q}=\left[\hat{\Sigma}^{-1}-\hat{\mathcal{G}}_0(0)\right]^{-1}$. Here $\hat{\mathcal{G}}_0$ is given by Eq.~(\ref{gsc_0}) and $\hat{\Sigma}$ is the self-energy associated with the internal structure of the impurity. From Eq.~(\ref{eq:ham_anderson}) we find that this self-energy is given by
\begin{equation}
\hat{\Sigma}=\frac{t_0^2\left(z\hat{\tau}_0+\epsilon_0\hat{\tau}_3 \right)}{z^2-\epsilon_0^2}. 
\end{equation}
Using these expressions the $T$-matrix for the quantum impurity takes the form:
\begin{equation}
\hat{T}_Q=t_0^2\frac{z\left(1-t_0^2g_0(0)\right) \hat{\tau}_0+\Delta t_0^2g_0(0)\hat{\tau}_1 + \left[\epsilon_0+t_0^2 g_3(0)\right]\hat{\tau}_3}{z^2\left[1-t_0^2 g_0(0)\right]^2-\Delta^2t_0^4g_0^2(0)-\left[\epsilon_0+t_0^2 g_3(0)\right]^2}.
\label{eq:tmatrix_imp}
\end{equation}
Combining these expressions, we find the exact Green's function for the superconductor in the presence of a quantum impurity:
\begin{equation}
\begin{aligned}
\hat{\mathcal{G}}(x_1,x_2)&=z \hat{\tau}_0 \left[g_0(x_1-x_2) +\delta g^{(Q)}_0(x_1,x_2)\right] \\
&+\Delta\hat{\tau}_1\left[g_0(x_1-x_2) +\delta f^{(Q)}_1(x_1,x_2) \right] \\
&+ \hat{\tau}_3\left[ g_3(x_1-x_2) + \delta g^{(Q)}_3(x_1,x_2) \right]  \\
&+ iz\Delta  \hat{\tau}_2  \delta f^{(Q)}_2(x_1,x_2), 
\end{aligned}
\label{eq:g_quant_imp}
\end{equation}
with coefficients $\delta g^{(Q)}_0$, $\delta g^{(Q)}_3$, $\delta f^{(Q)}_1$, and $\delta f^{(Q)}_2$ given by Eqs. (\ref{eq:delta_g_quant_imp}) in the Appendix. 
While we do not present the full expressions for these coefficients here, it is important to note that all four are even functions of frequency, $z$. Furthermore, while $\delta g^{(Q)}_0$, $\delta g^{(Q)}_3$, and $\delta f^{(Q)}_1$ are even under the exchange of the spatial coordinates $(x_1\leftrightarrow x_2)$, $\delta f^{(Q)}_2$ is odd under spatial coordinate exchange. Therefore, we directly find that the presence of a quantum impurity induces a pair amplitude with novel symmetry proportional to $\hat{\tau}_2$. Upon closer inspection, we see that this novel pair amplitude is directly proportional to the frequency, $z$, and, hence, it is an odd-$\omega$ odd-parity amplitude. Moreover, because this amplitude retains the spin-singlet nature of the original condensate, we conclude that it belongs to the symmetry class in the sixth column of Table \ref{table:classification}: odd-$\omega$, spin-singlet, and odd-parity. 

It is curious that the even- and odd-$\omega$ pair amplitudes in Eq. (\ref{eq:g_quant_imp}) correspond precisely to the coefficients of $\hat{\tau}_1$ and $\hat{\tau}_2$, respectively, when their symmetries only differ in the spatial and frequency domains and the $\hat{\tau}_i$ matrices are not obviously related to either of these degrees of freedom. However, we gain insight by noting that the reason there is no $\hat{\tau}_2$ term in the bare superconducting Green's function, Eq.~(\ref{gsc_0}), is that $\Delta\in \mathbb{R}$, which is a choice we can always make for a uniform superconducting system. In fact, if instead $\Delta=|\Delta|e^{i\phi}$, then a term proportional to $|\Delta|\sin{\phi}\hat{\tau}_2$ would appear in Eq.~(\ref{gsc_0}). Therefore, the odd-$\omega$ amplitude is proportional to $\hat{\tau}_2$ because, in the Matsubara representation, the even-$\omega$ pair amplitudes describe the real part of $F$, while the odd-$\omega$ terms describe the imaginary part of $F$. The latter can be understood as a direct consequence of the fact that the odd-$\omega$ terms must be imaginary to preserve time-reversal symmetry.

%This decomposition can easily be proved if we assume that the imaginary time propagator $F(\tau)$, as defined in Eq.~(\ref{eq:g_f_sc}) is real-valued, which is consistent with our assumption of $\Delta\in \mathbb{R}$. This assumption is valid for this system because we are able to choose any value of $\phi$, here, by performing a suitable $U(1)$ transformation.}

While the precise expression for the odd-$\omega$ pair amplitude for a quantum impurity, $F_{\text{odd},Q}=z\Delta\delta f^{(Q)}_{2}$, is a bit cumbersome, in the limit of weak coupling between the impurity level and the substrate, $t_0^2 \ll v_F$, and assuming $\Delta \ll \mu$, which generally holds for BCS superconductors, it simplifies considerably and we have
\begin{equation}
F_{\text{odd},Q}(x_1,x_2)=-\frac{iz\Delta t_0^2 \sin\left[\zeta k_F\left( |\tilde{x}_1|-|\tilde{x}_2|\right)\right]}{v_F^2\Omega\left(z^2-\epsilon_0^2 \right)\exp\left[-i\zeta\Omega \tfrac{|\tilde{x}_1|+|\tilde{x}_2|}{v_F}\right]},
\label{eq:fodd_quant}
\end{equation}
where $\tilde{x}_i=x_i-x_0$ is the $x$-coordinate measured from the position of the impurity, $x_0$. From this expression we see that the odd-$\omega$ amplitude exhibits a sinusoidal oscillation in the spatial separation, $|\tilde{x}_1|-|\tilde{x}_2|$, with period $2\pi/k_F$. At low frequencies, i.e.~$z \ll \Delta$, this amplitude also decays exponentially when the average distance from the impurity, $|\tilde{x}_1|+|\tilde{x}_2|$, is much greater than the superconducting coherence length, $\xi= v_F/\Delta$. Additionally, we note that this amplitude has acquired new poles at $z=\pm \epsilon_0$, due to the propagator associated with the impurity level. Therefore, the size of this odd-$\omega$ pair amplitude is strongly enhanced at $z=\pm\epsilon_0$ for $|\epsilon_0|<|\Delta|$ but becomes suppressed for $|\epsilon_0|>>|\Delta|$. This further emphasizes the point that the internal structure of the impurity state is important in determining the magnitude of the induced odd-$\omega$ pairing.

From the results of this section, it is clear that the presence or absence of internal structure in the non-magnetic impurity determine whether or not odd-$\omega$ pairing is induced. This shows that an internal structure to an impurity or otherwise translation symmetry breaking setup, such as at an SN interface, generates odd-$\omega$ pairing.
However, so far we have also assumed that the order parameter in the superconductor is homogeneous, an assumption we will next relax. 

\section{Pair Symmetry in the presence of an anomalous impurity}
\label{sec:anomalous_imp}
While Anderson's theorem assures us that conventional $s$-wave superconductors are insensitive to non-magnetic disorder,\cite{anderson1959theory} non-magnetic impurities can still modify the properties of clean $s$-wave superconductors through local variations of the gap, proportional to the impurity strength.\cite{fetter1965spherical,zhitomirsky1998effect,lauke2018friedel} Moreover, drastic changes of the superconducting gap take place at the interface of SN junctions. Therefore, we here consider the effect of a spatially inhomogeneous gap using an anomalous impurity, which locally correct the superconducting gap.

To capture the qualitative features of a spatially inhomogeneous gap, we consider a model for a clean 1D superconductor in the presence of a local correction to $\Delta$:       
\begin{equation}
H=H_{\text{SC}}+H_{\delta\Delta}
\label{eq:ham_anom}
\end{equation}
where $H_{\text{SC}}$ is the Hamiltonian describing the clean 1D superconductor, given by Eq.~(\ref{eq:ham_sc}), and  $H_{\delta\Delta}$ is a local perturbation of the superconducting order parameter described by
\begin{equation}
\begin{aligned}
H_{\delta\Delta}&=\delta\Delta\int dx \delta(x-x_0) \psi^\dagger_{x,\uparrow}\psi^\dagger_{x,\downarrow} + \text{H.c.}, 
\end{aligned}
\label{eq:ham_pot_gap}
\end{equation}
where $\delta\Delta$ parameterizes the local deviation from the clean gap, $\Delta$. In this case, the total Green's function is given by Eq.~(\ref{eq:g_pot}) but with the $T$-matrix
\begin{equation}
\hat{T}_{\delta\Delta}=-\delta\Delta\frac{\left( z  \hat{\tau}_0-\Delta\hat{\tau}_1 \right)\delta\Delta g_0(0)+\hat{\tau}_1 -\delta\Delta g_3(0)\hat{\tau}_3}{z^2\delta\Delta^2g_0^2(0)-\delta\Delta^2 g^2_3(0)-\left(1-\Delta\delta\Delta g_0(0) \right)^2}. 
\label{eq:T_dD}
\end{equation}
We therefore readily find that
\begin{equation}
\begin{aligned}
\hat{\mathcal{G}}(x_1,x_2)&=z \hat{\tau}_0 \left( g_0(x_1-x_2) +\delta g^{(\Delta)}_0(x_1,x_2) \right)  \\
&+ \Delta \hat{\tau}_1 \left( g_0(x_1-x_2) +\delta f^{(\Delta)}_1(x_1,x_2) \right)  \\
&+ \hat{\tau}_3\left( g_3(x_1-x_2) + \delta g^{(\Delta)}_3(x_1,x_2) \right)  \\
&+ iz \delta f^{(\Delta)}_2(x_1,x_2) \hat{\tau}_2  
\end{aligned}
\label{eq:g_delta_imp}
\end{equation}
with the coefficients $\delta g^{(\Delta)}_0$, $\delta g^{(\Delta)}_3$, $\delta f^{(\Delta)}_1$, and $\delta f^{(\Delta)}_2$ given by Eq.~(\ref{eq:delta_g_delta_imp}) in the Appendix. 

Similar to what we found in Sec.~\ref{sec:quantum_imp}, all four coefficients are even functions of the frequency, $z$, and $\delta g^{(\Delta)}_0$, $\delta g^{(\Delta)}_3$, and $\delta f^{(\Delta)}_1$ are even under the exchange of the spatial coordinates $(x_1\leftrightarrow x_2)$, while $\delta f^{(\Delta)}_2$ is odd under this coordinate exchange. Therefore, we find that the presence of an anomalous impurity gives rise to an odd-$\omega$ pair amplitude proportional to $\hat{\tau}_2$ and belonging to the symmetry class in the sixth column of Table \ref{table:classification}: odd-$\omega$, spin-singlet, and odd-parity. 

\begin{figure}
 \begin{center}
  \centering
        \includegraphics[width=0.45\textwidth]{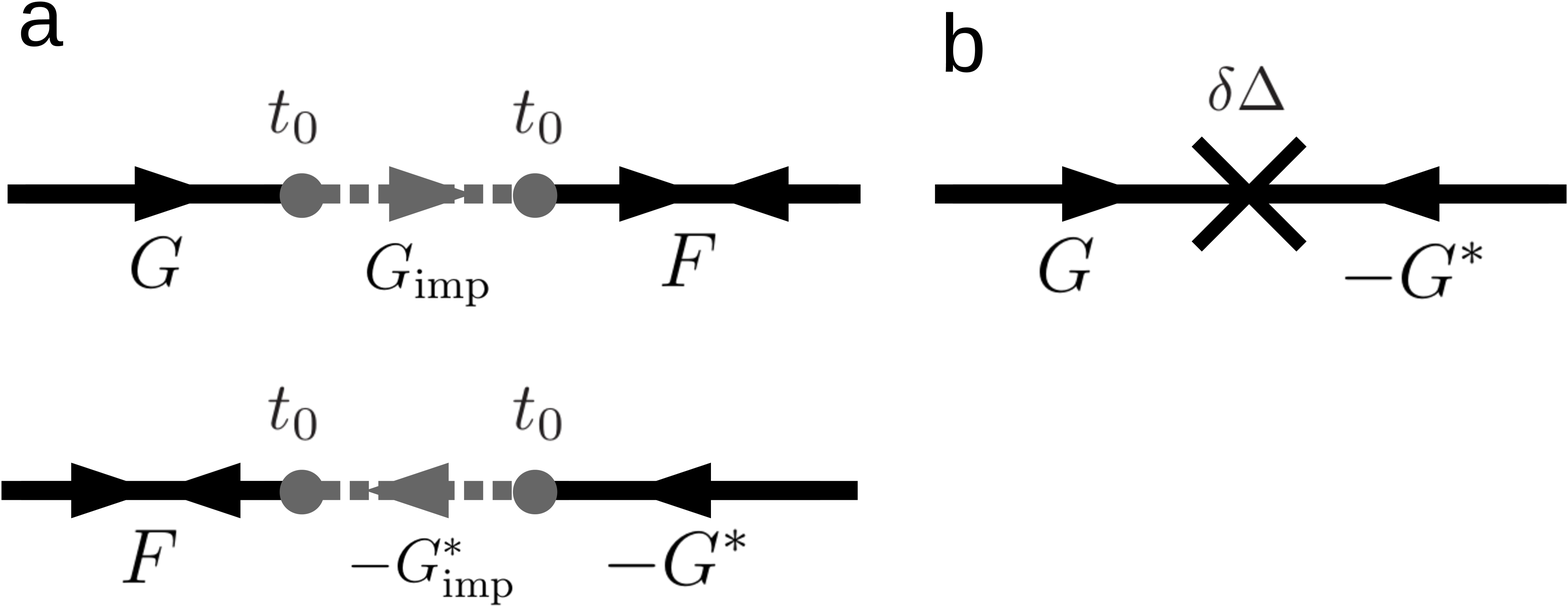}
  \caption{Diagrammatic representation of the leading order terms that give rise to the odd-$\omega$ pairing for a quantum impurity (a) and an anomalous impurity (b).}
  \label{fig:diagrams}
 \end{center}
\end{figure}

Turning our attention to the expressions for this odd-$\omega$ pair amplitude, we find that, similar to the case of the quantum impurity, they can be simplified considerably in the limit of $\Delta \ll \mu$, and assuming weak perturbations of the gap, $\delta\Delta \ll v_F$. In this limit we find
\begin{equation}
F_{\text{odd},\Delta}(x_1,x_2)=-\frac{iz\delta\Delta \sin\left[\zeta k_F\left( |\tilde{x}_1|-|\tilde{x}_2|\right)\right]}{v_F^2\Omega\exp\left[-i\zeta\Omega \tfrac{|\tilde{x}_1|+|\tilde{x}_2|}{v_F}\right]}.
\label{eq:fodd_delta}
\end{equation}
From this expression we see that this pair amplitude has exactly the same spatial dependence as the odd-$\omega$ amplitude in Eq.~(\ref{eq:fodd_quant}). However, there are important differences in their physical origin. These are most clearly illustrated by considering the leading order contributions to these amplitudes using diagrammatic perturbation theory, see Fig.~\ref{fig:diagrams}.
For the quantum impurity we find that, to leading order in the tunneling, $t_0$, the odd-$\omega$ pair amplitude is a result of two different scattering processes involving both quasiparticles and Cooper pairs, processes of the form $t_0^2G G_{\text{imp}} F$. This is illustrated in the two diagrams in Fig.~\ref{fig:diagrams}(a), where $G_{\text{imp}}$ is the normal state Green's function for the impurity.
In contrast, the odd-$\omega$ pairing induced by the anomalous impurity is generated purely by the scattering of normal quasiparticles by the local modification of the gap, $\delta\Delta$, that is, only processes of the form $G\delta\Delta G^*$ contribute, see Fig.~\ref{fig:diagrams}(b). Hence, the quantum impurity induces odd-$\omega$ pairing by allowing quasiparticles from the substrate to tunnel into the impurity state for a certain amount of time before rejoining the superconducting substrate as part of a Cooper pair, while in the case of the anomalous impurity the odd-$\omega$ pairing is caused by a ``gluing-together" of quasiparticles at different times by scattering processes involving the anomalous impurity field. 

The key role played by the impurity Green's function in the generation of odd-$\omega$ pair amplitudes for the quantum impurity case provides an explanation for why the potential impurity is not able to induce odd-$\omega$ pairing: the quasiparticles do not spend a finite amount of time on the classical impurity site. 
Still, the classical impurity will generate an odd-$\omega$ pair amplitude if it also changes the local superconducting gap. Since a classical impurity represents a local change in the DOS, which determines the magnitude of the order parameter, we expect the kind of odd-$\omega$ pairing appearing in Eq.~(\ref{eq:g_delta_imp}) to emerge in the presence of potential impurities. However, because this effect of the potential impurity is indirect, we expect these odd-$\omega$ pair amplitudes are only relevant in the presence of very strong potential impurities. 
The exception is clearly the SN interface where $\Delta$ necessarily changes to a zero value in the N part of the junction. We thus conclude that, not only can internal dynamics of the N region generate odd-$\omega$ pairing, but also the abrupt change of the superconducting order parameter, $\Delta$, can induce odd-$\omega$ pairing at SN interfaces.

We next explore how the fundamentally different physical origins for the odd-$\omega$ pair amplitudes give rise to qualitatively different dynamics.

\section{Comparison of Odd-Frequency Pairing in the time domain}
\label{sec:time}

To obtain deeper insight into the physical origins of the different odd-$\omega$ pair amplitudes emerging in the presence of the quantum impurity found in Sec.~\ref{sec:quantum_imp} and those generated by the anomalous impurity considered in Sec.~\ref{sec:anomalous_imp}, we Fourier transform the $\omega$-dependent results to the time domain. In general, the time-ordered Green's function, $G^{T}(t)$, is given by:
\begin{equation}
G^{T}(t)=\int\frac{d\omega}{2\pi}e^{-it\omega}\left[\theta(\omega)G^{R}(\omega)+\theta(-\omega)G^{A}(\omega) \right],
\end{equation} 
where $G^{R}(\omega)$ ($G^{A}(\omega)$) is the retarded (advanced) Green's function in the frequency domain. 
In both Eqs.~(\ref{eq:fodd_quant}) and (\ref{eq:fodd_delta}), we have expressed the odd-$\omega$ pair amplitude in terms of the complex frequency. From these expressions, we obtain the retarded and advanced Green's functions by taking $z=\omega + i0^{+}$ and $z=\omega + i0^{-}$, respectively. Therefore, in both cases, we can write the time-ordered odd-$\omega$ pair amplitude in the time-domain as:
\begin{equation}
\begin{aligned}
\tilde{F}_{odd}(t)&=\int_{-\infty}^{\infty}\frac{d\omega}{2\pi}e^{-it\omega}\left[\theta(\omega)F_{odd}(\omega+i0^+) \right. \\
&+\left.\theta(-\omega)F_{odd}(\omega+i0^-) \right]. \\
\end{aligned}
\label{eq:odd_time_domain_general}
\end{equation}   

Inserting the expressions for the odd-$\omega$ pair amplitudes from Eqs.~(\ref{eq:fodd_quant}) and (\ref{eq:fodd_delta}) into Eq.~(\ref{eq:odd_time_domain_general}), we obtain the following two expressions for the odd-$\omega$ amplitudes in the time-domain:
\begin{equation}
\begin{aligned}
\tilde{F}_{odd,Q}(t)&=-2t_0^2\Delta \ v_F^{-2}\sin\left[k_F\left(|\tilde{x}_1|-|\tilde{x}_2|\right)\right]f_Q(t), \\
\tilde{F}_{odd,\Delta}(t)&=-2\delta\Delta \ v_F^{-2}\sin\left[k_F\left(|\tilde{x}_1|-|\tilde{x}_2|\right)\right]f_\Delta(t),
\end{aligned}
\label{eq:fodd_t1}
\end{equation}
with
\begin{equation}
\begin{aligned}
f_{Q}(t)&=\int_0^{\infty}\frac{d\omega}{2\pi}\frac{\sin(\omega t)(\omega+i0^+)e^{i\tfrac{d}{v_F}\sqrt{(\omega+i0^+)^2-\Delta^2}}}{\sqrt{(\omega+i0^+)^2-\Delta^2}\left[(\omega+i0^+)^2-\epsilon_0^2\right]}, \\
f_{\Delta}(t)&=\int_0^{\infty}\frac{d\omega}{2\pi}\frac{\sin(\omega t)(\omega+i0^+)e^{i\tfrac{d}{v_F}\sqrt{(\omega+i0^+)^2-\Delta^2}}}{\sqrt{(\omega+i0^+)^2-\Delta^2}},
\end{aligned}
\label{eq:f_t}
\end{equation}
where $d = |\tilde{x}_1|+|\tilde{x}_2|$. Notice that both odd-$\omega$ amplitudes in Eq.~(\ref{eq:fodd_t1}) have essentially the same form, the only substantial difference is in the time-dependent factors in Eq.~(\ref{eq:f_t}). Comparing these two factors we see that the difference is that the integrand of $f_{Q}$ possesses additional poles in frequency-space, implying that, in the time domain, we expect $f_{Q}(t)$ to be much broader than $f_{\Delta}(t)$.
To better understand the behavior, it is convenient to restrict ourselves to positive times, $t>0$, and timescales short relative to $\tau_\xi=\xi/v_F=1/\Delta$. In this case we find 
\begin{equation}
\begin{aligned}
f_{Q}(t)&=t\int_0^{\infty}\frac{du}{2\pi}\frac{\sin(u)e^{i\tfrac{d}{v_Ft}u}}{(u+i0^+)^2-(t\epsilon_0)^2}, \\
f_{\Delta}(t)&=\frac{t-e^{i\tfrac{d}{v_F}\Lambda}\left[t\cos\left(t\Lambda\right)-i\tfrac{d}{v_F}\sin\left(t\Lambda\right) \right]}{2\pi\left[t^2-\left(\tfrac{d}{v_F}\right)^2 \right]},
\end{aligned}
\label{eq:f_t_simple}
\end{equation}
where we replaced the upper limit of the integral in $f_\Delta(t)$ with a finite cutoff, $\Lambda$, to obtain a convergent expression.

In the case of the quantum impurity, we see that at $t \ll 1/\epsilon_0$, the odd-$\omega$ amplitude increases as $\sim t$, while for $t \gg 1/\epsilon_0$ the amplitude decreases as $\sim t^{-1}$. This non-local behavior in time is a direct consequence of the dynamics of the impurity itself. Physically, this temporal broadening is an indication that the Cooper pair electrons tunnel onto the impurity site, where they remain for some duration, before tunneling back to the superconductor. This gives rise to the unequal-time correlations necessary to support odd-$\omega$ pairing.   

In contrast to the temporally broad behavior of the quantum impurity, we see that for the anomalous impurity the odd-$\omega$ amplitude is highly localized in the time domain, as evidenced by the strong peak at $t=d/v_F=(|x_1-x_0|+|x_2-x_0|)/v_F$ in Eq.~(\ref{eq:f_t_simple}). This time scale is precisely the amount of time it takes a quasiparticle to propagate from point $x_1$ to the impurity at $x_0$ and then from the impurity to the point $x_2$. Hence, this odd-$\omega$ pair amplitude represents Cooper pairs formed from normal quasiparticle states at $x_1$ and $x_2$ bound together by scattering off of the local gap perturbation, $\delta\Delta$, at $x_0$, in complete agreement with our previous perturbation theory findings.  

\section{Conclusions}
\label{sec:conclusions}
Within the literature it is often found that breaking a symmetry of a conventional superconductor gives rise to odd-$\omega$ pair amplitudes proportional to the symmetry breaking field.\cite{BergeretPRL2001, bergeret2005odd, halterman2007odd, yokoyama2007manifestation, houzet2008ferromagnetic, EschrigNat2008, LinderPRB2008, crepin2015odd, YokoyamaPRB2012, Black-SchafferPRB2012, Black-SchafferPRB2013, TriolaPRB2014, tanaka2007theory, TanakaPRB2007,cayao2017odd, cayao2018odd, LinderPRL2009, LinderPRB2010_2, TanakaJPSJ2012, triola2016prl, triolaprb2016, black2013odd, sothmann2014unconventional, parhizgar_2014_prb, asano2015odd, komendova2015experimentally, burset2016all, komendova2017odd, kuzmanovski2017multiple, triola2017pair,keidel2018tunable, triola2018odd,fleckenstein2018conductance,asano2018green,triola2018oddnw} 
One of the simplest examples of this phenomenon is the generation of odd-$\omega$ odd-parity pairing near the interface of superconductor-normal metal (SN) junctions, assumed to be related to the translation-symmetry breaking of the SN interface.\cite{tanaka2007theory, TanakaPRB2007} 
In this work we provided the microscopic mechanism behind the generation of this kind of odd-$\omega$ pairing. Specifically, we considered one of the simplest setups where translation symmetry is broken: a uniform one-dimensional $s$-wave spin-singlet superconductor with a single non-magnetic impurity localized at position $x_0$.
Surprisingly, by examining the emergent pair symmetries due to the presence of the impurity, we found that translation-symmetry breaking is actually not enough to induce odd-$\omega$ pair amplitudes. 

To understand how odd-$\omega$ pairing emerges, we modeled the impurity in three different ways: (i) as a classical potential impurity; (ii) as a quantum impurity with an on-site energy level coupled to the superconductor through a tunneling amplitude; and (iii) as an anomalous impurity modifying the local superconducting gap. Using these models we demonstrated that the potential impurity does not induce odd-$\omega$ pairing, despite the fact that it manifestly breaks translation invariance. Moreover, no amount of potential impurities can directly induce odd-$\omega$ pairing. However, we found that both the quantum impurity and the anomalous impurity do induce odd-$\omega$ pairing, but with very different physical origins. 

The odd-$\omega$ amplitude induced by the quantum impurity emerges when single electrons from the condensate tunnel to the impurity site and then tunnel back to the superconductor at a later time. In contrast, the odd-$\omega$ amplitude induced by the anomalous impurity is a direct result of the scattering of two normal quasiparticles from the superconductor by the local perturbation of the gap, essentially ``gluing together" two electrons from different positions and times. Fourier-transforming these different amplitudes to the time-domain, we found that the separate origins lead to qualitatively different temporal profiles. The quantum impurity produces odd-$\omega$ pairing with a temporal broadening inversely proportional to the energy level of the impurity, while the anomalous impurity leads to odd-$\omega$ pairing that is a highly localized in the relative time coordinate.

For a simple overview we summarize the main results in Table~\ref{table:results}.
We stress that, while these results were obtained for a one-dimensional model, the extension of these results to higher dimensions would only change the precise functional forms of the coefficents $g_0$ and $g_3$ in Eq. (\ref{gsc_0}). Importantly, the matrix structure in Eq. (\ref{gsc_0}), and both the frequency and spatial parities of $g_0$ and $g_3$ will be identical in all higher dimensions. Therefore, our conclusions about the presence or absence of odd-$\omega$ pairing, in each of these cases, do not depend on the dimensionality of the problem.

\begin{center}
\begin{table}[htb]
\begin{tabular}{|c | c | c | c |}
\hline
Impurity & $\hat{V}$ & Odd-$\omega$ Pairing & $t$-dependence  \\
\hline 
Potential  & $\delta\mu\hat{\tau}_3$ & none & NA  \\ 
\hline
Quantum  & $\tfrac{t_0^2}{z^2-\epsilon_0^2}\left(z\hat{\tau}_0+\epsilon_0\hat{\tau}_3 \right)$ & yes & non-local in $t$ \\
\hline
Anomalous & $\delta\Delta\hat{\tau}_1$ & yes & local in $t$ \\
\hline
\end{tabular}
\caption{Summary of main results for the three impurities considered in this work, indicated in left column. In each case the form of the scattering is specified using $\hat{V}$, which appears in the $T$-matrix, defined as: $\hat{T}=[\hat{V}^{-1}-\hat{\mathcal{G}}_0 ]^{-1}$.}
\label{table:results}
\end{table}
\end{center}

In summary, these results demonstrate that translation-symmetry breaking is by itself not sufficient for inducing odd-$\omega$ pairing in a conventional $s$-wave superconductor, but some additional structure is needed. This additional structure can be in the form of a quantum level, allowing the generation of substantial unequal-time correlations in the superconductor. Alternatively, a spatially inhomogeneous order parameter can generate odd-$\omega$ pairing. 
Thus this work not only provides possible microscopic mechanisms for generating odd-$\omega$ pairing in non-magnetic systems and establishes the resulting behavior of the odd-$\omega$ pair correlations, but importantly also opens the door to designing systems with stronger and targeted odd-$\omega$ pairing behavior.

\acknowledgments 
We thank A.~V.~Balatsky, J.~Cayao, J.~Fransson, T.~L\"{o}thman, M.~Mashkoori, and F.~Parhizgar for useful discussions. This work was supported by the Swedish Research Council (Vetenskapsr\aa det) Grant No. 621-2014-3721 and 2018-03488, the Knut and Alice Wallenberg Foundation through the Wallenberg Academy Fellows program, and the European Research Council (ERC) under the European Unions Horizon 2020 research and innovation programme (ERC-2017-StG-757553).

\appendix 

\section{Green's function coefficients}
\label{app:coeffs}

Using the $T$-matrix in Eq.~(\ref{eq:tmatrix_imp}), we find the Green's function for the superconductor in the presence of a quantum impurity, given by Eq.~(\ref{eq:g_quant_imp}), with coefficients: 
\begin{equation}
\begin{aligned}
\delta g^{(Q)}_0&(x_1,x_2)=\frac{t_0^4g_0(0)}{D_Q}\left\{  \left(\tfrac{1}{t_0^2g_0(0)}-1\right) g_3(\tilde{x}_1) g_3(\tilde{x}_2) \right. \\
&+\left.  \tfrac{\epsilon_0+t_0^2g_3(0)}{t_0^2g_0(0)}\left[g_0(\tilde{x}_1) g_3(\tilde{x}_2)+g_3(\tilde{x}_1) g_0(\tilde{x}_2) \right]  \right.\\
&-\left. \left(\Omega^2-\tfrac{z^2+\Delta^2}{t_0^2g_0(0)}\right)g_0(\tilde{x}_1) g_0(\tilde{x}_2)\right\}, \\
\delta g^{(Q)}_3&(x_1,x_2)=\frac{t_0^4g_0(0)}{D_Q}\left\{\tfrac{\epsilon_0+t_0^2g_3(0)}{t_0^2g_0(0)} \Omega^2 g_0(\tilde{x}_1) g_0(\tilde{x}_2) \right. \\
&-\left. \left( \Omega^2 -\tfrac{z^2}{t_0^2g_0(0)}\right) \left[g_0(\tilde{x}_1) g_3(\tilde{x}_2)+g_3(\tilde{x}_1) g_0(\tilde{x}_2) \right] \right. \\ 
&+\left.  \tfrac{\epsilon_0+t_0^2g_3(0)}{t_0^2g_0(0)}  g_3(\tilde{x}_1) g_3(\tilde{x}_2) \right\}, \\
\delta f^{(Q)}_1&(x_1,x_2)=\frac{t_0^4g_0(0)}{D_Q}\left\{ \left(\tfrac{2z^2}{t_0^2g_0(0)}-\Omega^2\right) g_0(\tilde{x}_1) g_0(\tilde{x}_2) \right. \\ 
&+\left.\tfrac{\epsilon_0+t_0^2g_3(0)}{t_0^2g_0(0)}\left[g_0(\tilde{x}_1) g_3(\tilde{x}_2)+g_3(\tilde{x}_1) g_0(\tilde{x}_2) \right] \right. \\
&- \left. g_3(\tilde{x}_1) g_3(\tilde{x}_2)\right\}, \\
\delta f^{(Q)}_2&(x_1,x_2)=\frac{t_0^2}{D_Q}\left\{g_3(\tilde{x}_1) g_0(\tilde{x}_2)-g_0(\tilde{x}_1) g_3(\tilde{x}_2)\right\}, 
\end{aligned},
\label{eq:delta_g_quant_imp}
\end{equation}
where 
\begin{equation}
D_Q=z^2\left[1-t_0^2 g_0(0)\right]^2-\Delta^2t_0^4g_0^2(0)-\left[\epsilon_0+t_0^2 g_3(0)\right]^2.
\end{equation}

Using the $T$-matrix in Eq.~(\ref{eq:T_dD}), we find the Green's function for the superconductor in the presence of an anomalous impurity, given by Eq.~(\ref{eq:g_delta_imp}), with coefficients:
\begin{equation}
\begin{aligned}
\delta g^{(\Delta)}_0&(x_1,x_2)=\frac{\delta\Delta^2}{D_\Delta}\left\{ g_3(0)\left[ g_0(\tilde{x}_1) g_3(\tilde{x}_2)+g_3(\tilde{x}_1) g_0(\tilde{x}_2)\right] \right. \\
&\left.-g_0(0)\left(\Omega^2+\tfrac{2\Delta}{\delta\Delta g_0(0)} \right)g_0(\tilde{x}_1) g_0(\tilde{x}_2) \right. \\
&\left.-g_0(0)g_3(\tilde{x}_1) g_3(\tilde{x}_2) \right\}, \\
\delta g^{(\Delta)}_3&(x_1,x_2)=\frac{\delta\Delta^2}{D_\Delta} \left\{g_3(0) \Omega^2 g_0(\tilde{x}_1) g_0(\tilde{x}_2)  \right. \\
&\left.-g_0(0)\left(\Omega^2+\tfrac{\Delta}{\delta\Delta g_0(0)}\right)  \left[ g_0(\tilde{x}_1) g_3(\tilde{x}_2)+g_3(\tilde{x}_1) g_0(\tilde{x}_2)\right] \right. \\
&\left.+ g_3(0) g_3(\tilde{x}_1) g_3(\tilde{x}_2)\right\}, \\
\delta f^{(\Delta)}_1&(x_1,x_2)=\frac{\delta\Delta^2}{D_\Delta} \left\{g_3(0)\left[ g_0(\tilde{x}_1) g_3(\tilde{x}_2)+g_3(\tilde{x}_1) g_0(\tilde{x}_2)\right] \right. \\
&\left.-g_0(0) \left(\Omega^2+\tfrac{z^2+\Delta^2}{\Delta \delta\Delta g_0(0)} \right) g_0(\tilde{x}_1) g_0(\tilde{x}_2) \right. \\
&\left.-g_0(0) \left(1-\tfrac{1}{\Delta \delta\Delta g_0(0)} \right)g_3(\tilde{x}_1) g_3(\tilde{x}_2)  \right\}, \\
\delta f^{(\Delta)}_2&(x_1,x_2)=\frac{\delta\Delta}{D_\Delta} \left\{ g_0(\tilde{x}_1) g_3(\tilde{x}_2)-g_3(\tilde{x}_1) g_0(\tilde{x}_2)\right\}, 
\end{aligned} 
\label{eq:delta_g_delta_imp}
\end{equation}
where
\begin{equation}
D_\Delta=z^2\delta\Delta^2g_0^2(0)-\delta\Delta^2 g^2_3(0)-\left(1-\Delta\delta\Delta g_0(0) \right)^2.
\end{equation}

\bibliographystyle{apsrevmy}
\bibliography{Odd_Frequency_Impurity}

\end{document}